# Incoherent magnetization dynamics in strain mediated switching of magnetostrictive nanomagnets

Dhritiman Bhattacharya[1, 2], Md Mamun Al-Rashid[1, 2], Noel D'Souza[1], Supriyo Bandyopadhyay[2], and Jayasimha Atulasimha[1, 2] *

[1]Department of Mechanical and Nuclear Engineering, Virginia Commonwealth University, Richmond, VA 23284, USA
[2]Department of Electrical and Computer Engineering, Virginia Commonwealth University, Richmond, VA 23284, USA

* Corresponding author: jatulasimha@vcu.edu

Micromagnetic studies of the magnetization change in magnetostrictive nanomagnets subjected to stress are performed for nanomagnets of different sizes. The interplay between demagnetization, exchange and stress anisotropy energies is used to explain the rich physics of size-dependent magnetization dynamics induced by modulating stress anisotropy in planar nanomagnets. These studies have important implications for strain mediated ultralow energy magnetization change in nanomagnets and its application in energy-efficient nanomagnetic computing systems.

## I. INTRODUCTION

The non-volatility and unprecedented energy efficiency of nanomagnetic logic and memory devices have spurred investigations of different types of magnetization reversal schemes in nanomagnets. The typical methods employed to induce magnetization reversal include the use of electric current-generated magnetic field [1], spin transfer torque [2], current-driven domain wall motion [3], current-induced spin-orbit torque or spin Hall effect [4-5], or strain generated by applying an electrical voltage to a multiferroic nanomagnet [6-11]. While the field and spin torque driven magnetization dynamics have been studied extensively, strain mediated magnetization dynamics is a relatively new concept which promises high energy efficiency in switching nanomagnetic devices [12-13] but has not been studied as extensively as the others.

The physics underlying these different switching mechanisms have pronounced differences, particularly the manner in which the energy landscape of the nanomagnet is modified by the switching agent (Fig 1). An external magnetic field reshapes the potential energy landscape of the nanomagnet in such a way that the magnetization aligns itself along the direction of the applied field. However, unlike a magnetic field, spin transfer torque (STT) is non-conservative in nature and hence cannot be modeled and explained within a potential energy framework. Spin-polarized current is utilized to transfer spin angular momenta to the resident spins in a nanomagnet, which reorient the magnetic moments in the nanomagnet resulting in magnetization reversal [14, 15]. A full 180° switching (or complete magnetic reversal) is achievable using any of these strategies. In contrast, the maximum possible rotation that can be induced in magnetostrictive nanomagnets by strain is 90°. There are ways around this; precisely timed application and withdrawal of stress [16], dipole coupling with a neighboring nanomagnet [17], or sequential application of stress along different directions (with the aid of carefully designed electrode configuration) [18] can bring about a complete 180° rotation.

Many theoretical studies of magnetization dynamics use the macrospin approximation in which the magnetization rotation is considered to be a collective and coherent rotation of spins. This approach, however, lacks the framework to study the intricate details of the micromagnetic configuration during the switching process. Extensive studies have been conducted to examine the effects of incoherent magnetization rotation under magnetic fields [19-22] or spin transfer torque [23-30]. However, incoherent magnetization rotation in shape anisotropic nanomagnets subjected to stress is yet to be rigorously addressed and analyzed. In this study, we have performed rigorous micromagnetic simulations to study the peculiarities of the incoherent magnetization dynamics induced by modulation of the stress anisotropy in a nanomagnet with applied strain.

## II. THEORY AND MICROMAGNETIC MODELING

In relatively small nanomagnets (~50 nm lateral dimensions), the magnetization is spatially uniform owing to exchange interaction and all the spins rotate more or less in unison (coherently) when the magnetization evolves from one state to another. However, in nanomagnets having larger dimensions, the magnetization states are predominantly non-uniform, particularly during the switching process. These size effects are investigated in elliptical disk-shaped nanomagnets of various dimensions, specifically for stress induced magnetization dynamics. Uniaxial compression/tension is generated along the major axis (easy axis) of the ellipse to rotate the magnetization from the easy toward the in-plane hard axis (minor axis of the ellipse). During this magnetization rotation, the interplay between the exchange, magnetostatic and stress anisotropy energies determines whether the magnetization rotation is coherent or not, and the final magnetic configuration that is stabilized.

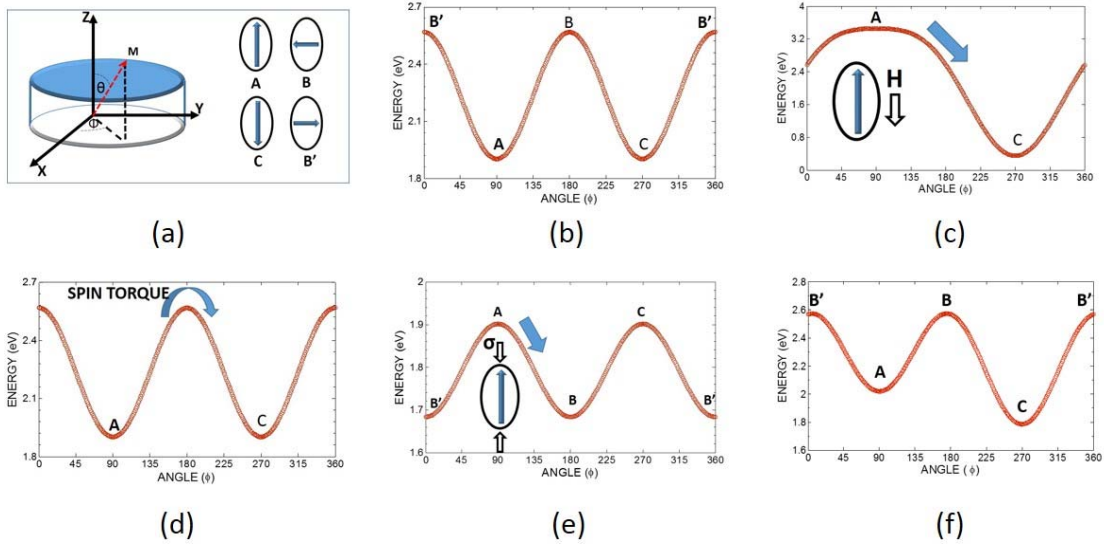

**Figure 1.** (a) The magnetization vector and four different states A, B, C and B' defined by the magnetization orientation in a nanomagnet shaped like an elliptical disk. (b) Energy profile of a shape anisotropic nanomagnet (shaped like an elliptical disk) as a function of magnetization orientation in the plane of the nanomagnet. The potential energy is plotted as a function of the azimuthal angle ϕ. (c) Modified energy profile with an applied magnetic field H, (d) spin transfer torque mediated switching from state A to state C, (e) modified energy profile with uniaxial compressive stress along the major axis for a magnet with positive magnetostriction (the same would be true for uniaxial tensile stress along the major axis for a magnet with negative magnetostriction), (f) introduction of asymmetry in the energy profile with a dipole coupled nanomagnet.

We performed micromagnetic simulations using the MuMax package [31]. Our geometry was discretized into $2\times2\times3$ nm$^3$ cells. Since Mumax does not have an inbuilt functionality to incorporate the effect of stress directly, we accomplish this through the use of uniaxial magnetocrystalline anisotropy, facilitated by their equivalent contribution to the effective field.

In the MuMax framework [31], the magnetization dynamics is simulated using the Landau Lifshitz Gilbert (LLG) equation:

$$\frac{\delta \vec{m}}{\delta t} = \vec{\tau} = (\frac{\gamma}{1+\alpha^2}) \times (\vec{m} \times \overrightarrow{H_{eff}} + \alpha \times (\vec{m} \times (\vec{m} \times \overrightarrow{H_{eff}}))) \quad (1)$$

where $m$ is the reduced magnetization (M/M$_s$), M$_s$ is the saturation magnetization, $\gamma$ is the gyromagnetic ratio and $\alpha$ is the Gilbert damping coefficient. The quantity $H_{eff}$ is the effective magnetic field which is given by,

$$\overline{H}_{eff} = \overline{H}_{demag} + \overline{H}_{exchange} + \overline{H}_{stress} \quad (2)$$

Here, H$_{demag}$, H$_{exchange}$ and H$_{stress}$ are the demagnetization (or magnetostatic) field, the effective field due to exchange coupling and effective field due to stress anisotropy evaluated in the MuMax framework in the manner described in Ref [31].

Table I lists the material parameters used in the simulation.

**TABLE I.** Material Parameters

| Parameters | Values |
| --- | --- |
| Exchange Stiffness Constant (A$_{ex}$) | $1.6\times10^{-11}$ J/m |
| Saturation Magnetization (M$_s$) | $1.32\times10^{6}$ A/m |
| Gilbert Damping Constant ($\alpha$) | 0.017 |
| Magnetostrictive Coefficient (3/2)($\lambda_s$) | $4\times10^{-4}$ |

### III. RESULTS

#### A. Pre-stress equilibrium states

For an ideal single domain elliptical nanomagnet, the in-plane hard and easy axes for the magnetization lie along the minor and major axis of the elliptical disk, respectively. The energy difference between the two states, i.e. when the magnetization is along the hard axis and when the magnetization is along the easy axis, is the in-plane energy barrier height, denoted by $\Delta E_1$ in Table II. While choosing the different nanomagnet dimensions to study the effect of stress, the aspect ratios (ratio of the major axis to the minor axis to the thickness) were held constant which, in turn, keeps the demagnetization factors identical over all chosen dimensions. In other words, all magnets that have been studied have approximately the same value of the critical stress ($\sigma_c$) under the macrospin assumption, where critical stress is defined as the stress required to make the stress anisotropy energy equal to the in-plane barrier height. The barrier height and corresponding critical stress values calculated for the chosen nanomagnet dimensions are shown in Table II. Note that the slight differences in $\sigma_c$ originate from the use of identical discretization volumes for all nanomagnet geometries that results in a relatively better staircase approximation in the larger nanomagnets.

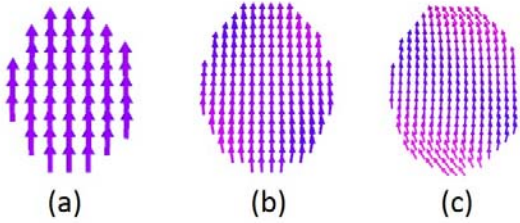

**Figure 2.** Equilibrium (relaxed) states of different sized nanomagnets showing spatial variation in the magnetization orientation (a) 50 nm×40 nm×3 nm, (b) 100 nm×80 nm×6 nm, (c) 200 nm×160 nm×12 nm.

In order to observe the size effects, the initial magnetization was set along the major axis, following which the nanomagnet's spins were allowed to relax. These relaxed or equilibrium states are shown in Fig. 2. The barrier height ($\Delta E_1^*$) and the critical stress ($\sigma_c^*$) were recalculated for the newly found easy states and increase with increasing dimensions (see Table II). The energy barrier per unit volume ($\Delta E_1/V$) between the states uniformly magnetized along the minor axis and those uniformly magnetized along the major axis are independent of the nanomagnet volume. However, the difference in energy per unit volume between the states uniformly magnetized along the major axis and the relaxed states ($\Delta E_{relaxed}/V$) increases with increasing volume. Thus, the energy difference per unit volume between the states uniformly magnetized along the hard axis and the "relaxed" states (which is essentially $\Delta E^*_1/V = \Delta E_1/V + \Delta E_{relaxed}/V$) also increases with the size of the nanomagnet. This leads to increasing $\sigma_c^*$ with increasing nanomagnet volume, where $\sigma_c^*$ is the stress at which the stress anisotropy energy equals $\Delta E^*_1$.

**TABLE II.** Barrier height and critical stress calculation for different sized magnets in single domain and micromagnetic scenario

| Dimension (nm×nm×nm) | $\Delta E_1$ (eV) | ($\sigma_c$) (MPa) | $\Delta E^*_1$ (eV) | ($\sigma_c^*$) (MPa) |
|---|---|---|---|---|
| 50×40×3 | 0.665 | 56.48 | 0.676 | 57.41 |
| 100×80×6 | 5.570 | 59.10 | 6.208 | 65.87 |
| 200×160×12 | 46.008 | 61.14 | 57.607 | 75.84 |

### B. Magnetization rotation under stress

First, the nanomagnet with the smallest dimensions (50 nm × 40 nm × 3 nm) is examined. Uniaxial stress was applied along the major axis and ramped up in discrete steps of 15 MPa. After each step, the magnetization was allowed to settle to the intermediate equilibrium state. Two intermediate stress points (the critical stress $\sigma_c^*$ and the stress of which rotates the magnetization by exactly 90°) were also added to the set to provide a complete picture. Ideally, if the process was entirely coherent, then these two stress values would be the same, i.e. the critical stress would abruptly rotate the magnetization from the major to the minor axis. Any difference between these two values is a measure of the incoherency. As can be seen in Fig. 3, the magnetization rotates in a predominantly coherent manner under stress (Fig. 3a, b) and aligns along the minor axis (~90° rotation) at a stress of 63 MPa (Fig. 3c) while the critical stress value shown in Table II is 57.41 MPa. The small difference is due to the minor incoherency of the rotation process. Upon removal of this stress, the magnetization has an equal probability of either rotating back to its initial orientation or flipping by 180°. This magnetization rotation picture most closely matches the macrospin (coherent rotation) approximation.

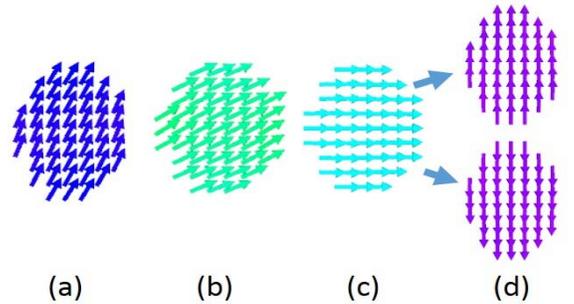

**Figure 3**. Evolution of magnetization with stress for a nanomagnet with dimensions of 50 nm × 40 nm × 3 nm, for the following stress scenarios: (a) 45 MPa, (b) 57.41 MPa (critical stress, $\sigma_c^*$) (c) 63 MPa (stress for 90º magnetization rotation), (d) Removal of stress.

In nanomagnets of intermediate dimensions (100 nm × 80 nm × 6 nm), the magnetization goes through a 'C-state' that is stable under stress application (Fig. 4). The 'C-state' is so-named because the spins curl into the shape of the letter 'C'. Stress was varied in increments of 15 MPa from 0 MPa up to a maximum of 180 MPa. As stress increases, the magnetization component along the minor axis gradually increases, but there still exists a component of the magnetization along the major axis (Fig. 4c). In other words, even a stress as high as 180 MPa does not rotate or modify the magnetization orientation so much that the memory of the prior state is completely erased. Thus, upon removal of stress, the magnetization always returns to its initial state (Fig. 4d) because of the retained memory. Also, because the magnetization reverts to the original state upon withdrawal of stress, the magnetization change is 'reversible' and the state visited under stress is 'volatile'.

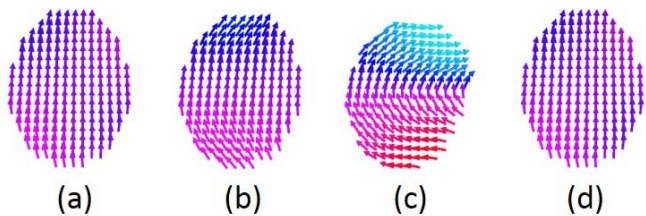

**Figure 4**. Evolution of magnetization with stress for a nanomagnet with dimensions of 100 nm × 80 nm × 6 nm, for the following stress scenarios: (a) 30 MPa, (b) 60 MPa, (c) 180 MPa, (e) Removal of stress.

Interestingly, in larger sized nanomagnets (200 nm × 160 nm × 12 nm), the magnetization begins to rotate at lower stress (10 MPa) (Fig. 5a), compared to the smaller and intermediate sized magnets, in which no significant magnetization rotation is observed until a stress of 45 MPa is applied. The magnetization attains a very stable vortex state with stress as low as 20 MPa (Fig. 5b). Further increase in stress results in a small increase in the magnetization vectors oriented along the minor axis (± *x*-axis). The net magnetization, however, remains zero. The vortex state is a stable state and the nanomagnet remains in this state even if stress is withdrawn after the vortex formation, resulting in `non-volatility'. Therefore, in larger nanomagnets, the magnetization state is 'irreversible" and the state visited under stress is 'non-volatile' in stark contrast to the case for intermediate sized nanomagnets.

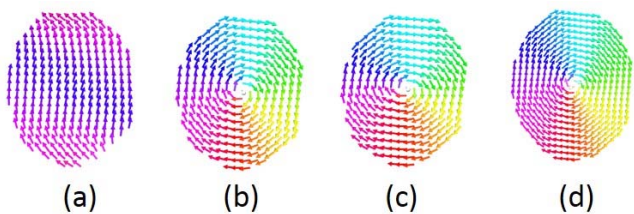

**Figure 5**. Evolution of magnetization with stress for a nanomagnet with dimensions of 200 nm × 160 nm × 12 nm, for the following stress scenarios: (a) 10 MPa, (b) 20 MPa, (c) 180 MPa, (d) Removal of stress.

### C. Study of the energy profile

The magnetization switching behavior of nanomagnets of different sizes can be explained by studying the energies involved. For this purpose, exchange, demagnetization, stress anisotropy and the total energies were calculated and plotted against applied stress.

For the smallest magnet, the demagnetization and stress anisotropy energies with all magnetization pointing exactly along the minor axis are shown by two reference lines (dashed lines in Fig. 6). Exchange energy is almost negligible, which explains the coherency of the switching process. There is an increase in demagnetization energy when the magnetization rotates from the major toward the minor axis. The magnetization orientation remains unchanged until a stress of 45 MPa is applied, since the demagnetization energy cost of rotating away from the easy axis is more than the lowering of stress anisotropy energy achieved by such a rotation. Thereafter, the magnetization starts to rotate, because the rate of increase of the demagnetization energy with increasing stress is less than the rate at which stress anisotropy decreases due to stress induced rotation (shown in Fig. 6). Therefore, the magnetization rotates to a slightly lower energy state. At 63 MPa stress, the stress anisotropy energy completely erodes the barrier between the easy and hard axis. Consequently, the minor axis orientation becomes energetically favorable and the magnet completes its magnetization rotation to this new easy state.

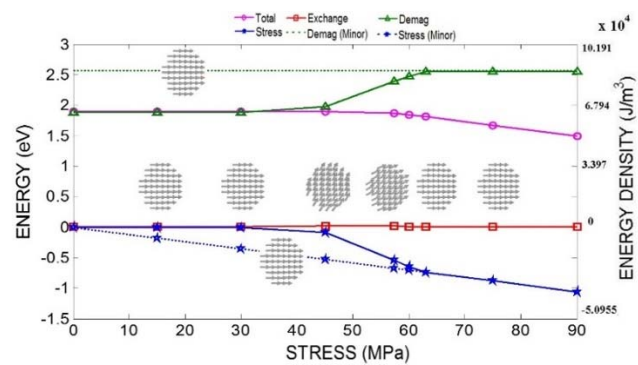

**Figure 6.** Total, exchange, demagnetization, stress anisotropy energies vs stress for a nanomagnet of dimensions 50 nm × 40 nm × 3 nm that show almost coherent magnetization rotation.

For the intermediate and large sized magnets, demagnetization and exchange energies of the vortex state are illustrated through two reference lines (dashed lines in Fig. 7 and Fig. 8). The calculation of total energy shows that vortex state is the energy minimum. In order to reach the vortex state from the initial equilibrium state, the nanomagnets need to go through intermediate states of high exchange energy. Hence, there exists a barrier between the vortex and initial equilibrium state which originates from the change in exchange energy it has to undergo to reach this vortex state. Therefore, the reduction in stress anisotropy energy and demagnetization energy should exceed the increase in exchange coupling energy to form the vortex state.

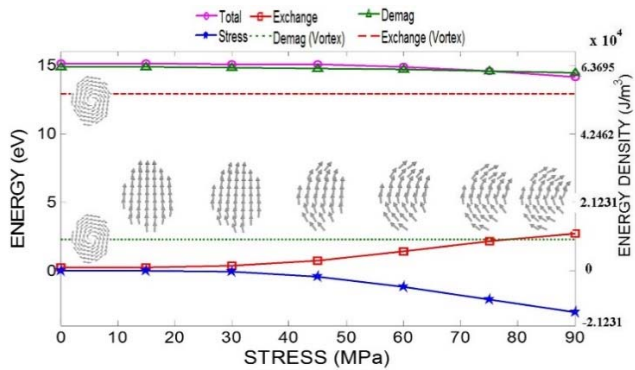

**Figure 7.** Total, exchange, demagnetization, stress anisotropy energies vs stress for a nanomagnet of dimensions 100 nm × 80 nm × 6 nm.

In the intermediate sized magnet, in the range of stress applied (~180 MPa, shown earlier in Fig 4) the magnet stabilizes itself in the C-shaped metastable state, although the vortex state is still the ultimate energy minimum state. Here the C-shaped persists up to large stresses as the exchange coupling penalty incurred in forming a vortex state is very high at the these lateral dimensions. On the other hand, the barrier between the initial equilibrium state and the vortex state is small for larger sized magnets. Therefore, at a stress of only 20 MPa, the magnetization settles to a vortex configuration in the larger sized nanomagnets (Fig. 8). Demagnetization and exchange energies of the nanomagnet meet the corresponding reference lines once the vortex is formed. Further increase in stress results in an increase in the component of the magnetization vector along the minor axis. Thus, a small increase in the demagnetization energy can be observed.

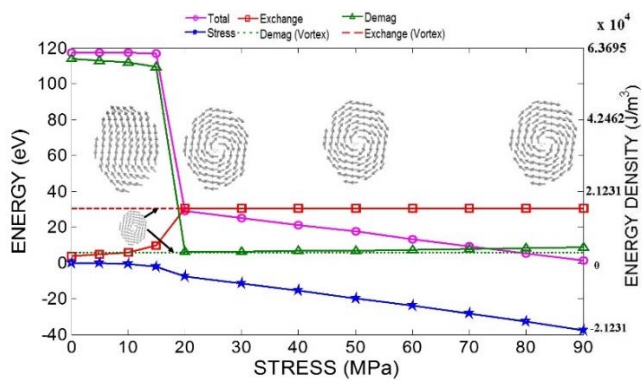

**Figure 8.** Total, exchange, demagnetization, stress anisotropy energies vs stress for a nanomagnet of dimensions 200 nm × 160 nm × 12 nm.

## IV. CONCLUSIONS

We have shown that the interplay between the exchange, demagnetization and stress energy can lead to a rich variety of magnetic configurations in planar shaped anisotropic nanomagnets under stress. These have been analyzed without including the effect of thermal noise to give us a simple physical picture. We discuss the magnetization behavior in the presence of thermal noise in the accompanying supplementary material. There is no qualitative change in the behavior in the presence of thermal noise for the two extreme cases: small nanomagnets with near coherent magnetization rotation and large nanomagnets where stress easily induces a stable vortex state. However, in the intermediate nanomagnet, in addition to the C-shape magnetization configuration formed in the absence of thermal noise, an intermediate configuration, where most of the magnetization points along the hard axis, also emerges when thermal noise is added.

This work can have implications for the use of nanomagnets switched with stress for low energy computing applications insofar as it shows that coherent rotation in small nanomagnets is perhaps the most viable switching strategy. If the vortex state is entered into (as in the large magnets), the field required to nudge the magnetization away from this vortex state is so high that it might negate any energy advantage of strain induced switching. Furthermore, these studies can shed light on some recent observations of irreversible magnetization change in large magnetostrictive nanomagnets [32].

## ACKNOWLEDGEMENTS

D.B, M.M.A. N.D and J.A are supported in part by the National Science Foundation CAREER grant CCF-1253370 and all authors are supported in part by the Commonwealth of Virginia, Center for Innovative Technology (CIT) managed Commonwealth Research Commercialization Fund (CRCF) Matching Funds Program.

# SUPPLEMENTARY INFORMATION

## Incoherent magnetization dynamics in strain mediated switching of magnetostrictive nanomagnets

Dhritiman Bhattacharya[1], Md Mamun Al-Rashid[1,2], Noel D'Souza[1], Supriyo Bandyopadhyay[2], and Jayasimha Atulasimha[1,2] *

[1]*Department of Mechanical and Nuclear Engineering, Virginia Commonwealth University, Richmond, VA 23284, USA*
[2]*Department of Electrical and Computer Engineering, Virginia Commonwealth University, Richmond, VA 23284, USA*

\* Corresponding author: jatulasimha@vcu.edu

In the main paper, we showed the key intermediate states that the magnetization of a nanomagnet subjected to stress goes through in the absence of thermal noise. We showed them for three different sizes of nanomagnets. Here we show a larger number of intermediate states that a nanomagnet goes through while subjected to stress in the absence of thermal noise. Additionally, to complete the picture, we also perform simulations in the presence of room-temperature thermal noise and study the effect of thermal perturbations on each of the above magnetization configurations. This provides a more realistic picture of magnetization dynamics at room temperature. The effect of thermal noise is modeled within the MuMax framework in the manner of reference [S1].

In the two extreme cases, i.e. for the smallest (Fig. S1 and S2) and the largest nanomagnet (Fig. S6 and S7), we found good qualitative agreement between the magnetic configurations in the presence and absence of thermal noise. However, the intermediate sized magnet shows a difference between the 0 K and 300 K results: It invariably reaches a C-state only in the absence of thermal noise (Fig S3), but when thermal noise is added, in addition to the C-state (Fig S4) it can reach another state as shown in Fig S5. Here, we observe an alternative to the C-state since thermal noise can produce sufficient deviation from the C-state to let the system evolve to a state in which most of the magnetizations rotate to point approximately parallel to the hard axis at large stresses. In other words, the original "relaxed" configuration biases the magnetization distribution in such a way that on application of stress, the easiest path for these magnetizations (spins) to minimize their energy is to rotate towards the C-state. However, thermal noise, enables the magnetization to overcome the effects of the initial bias, providing an alternate pathway to rotate under stress.

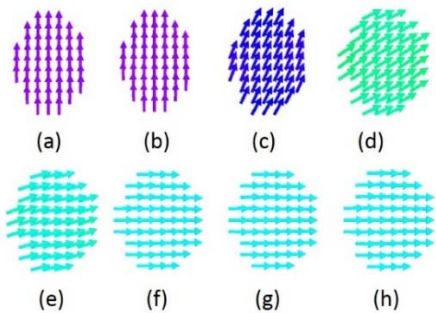

**Figure S1**: Evolution of magnetization at T=0 K with stress for a nanomagnet with dimensions of 50 nm × 40 nm × 3 nm, for the following stress scenarios: (a) 15 MPa, (b) 30 MPa, (c) 45 MPa, (d) 57.41 MPa, (e) 60 MPa, (f) 63 MPa, (g) 75 MPa, (h) 90 MPa.

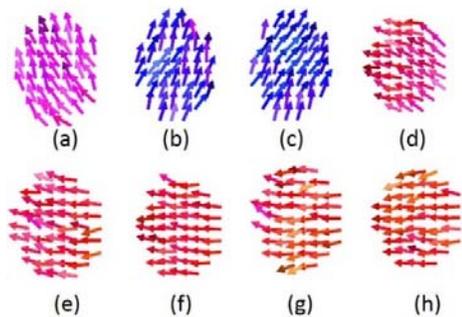

**Figure S2**: Evolution of magnetization at T=300 K with stress for a nanomagnet with dimensions of 50 nm × 40 nm × 3 nm, for the following stress scenarios: (a) 15 MPa, (b) 30 MPa, (c) 45 MPa, (d) 57.41 MPa, (e) 60 MPa, (f) 63 MPa, (g) 75 MPa, (h) 90 MPa. The spins are less coherent because of thermal noise.

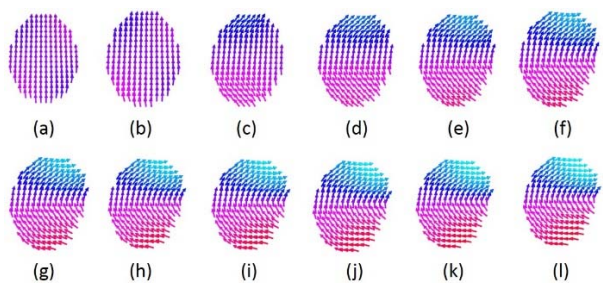

**Figure S3**: Evolution of magnetization at T=0 K with stress for a nanomagnet with dimensions of 100 nm × 80 nm × 6 nm, for the following stress scenarios: (a) 15 MPa, (b) 30 MPa, (c) 45 MPa, (d) 60 MPa, (e) 75 MPa, (f) 90 MPa, (g) 105 MPa, (h) 120 MPa, (i) 135 MPa, (j) 150 MPa, (k) 165 MPa, (l) 180 MPa.

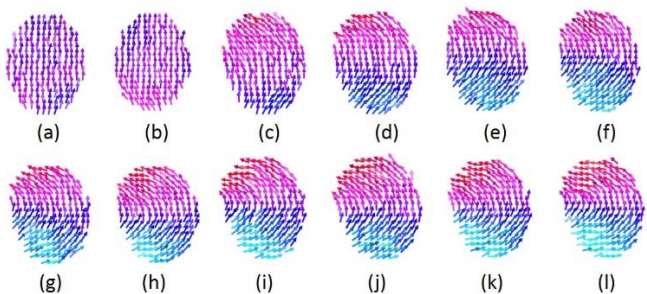

**Figure S4**: Evolution of magnetization at T=300 K with stress for a nanomagnet with dimensions of 100 nm × 80 nm × 6 nm, for the following stress scenarios: (a) 15 MPa, (b) 30 MPa, (c) 45 MPa, (d) 60 MPa, (e) 75 MPa, (f) 90 MPa, (g) 105 MPa, (h) 120 MPa, (i) 135 MPa, (j) 150 MPa, (k) 165 MPa, (l) 180 MPa. Once again, the magnetization shows more deviation from the C-state in Fig. 3 due to thermal noise.

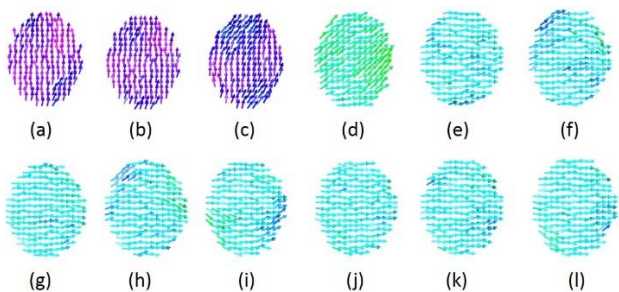

**Figure S5**: Evolution of magnetization at T=300 K with stress for a nanomagnet with dimensions of 100 nm × 80 nm × 6 nm, for the following stress scenarios: (a) 15 MPa, (b) 30 MPa, (c) 45 MPa, (d) 60 MPa, (e) 75 MPa, (f) 90 MPa, (g) 105 MPa, (h) 120 MPa, (i) 135 MPa, (j) 150 MPa, (k) 165 MPa, (l) 180 MPa. In this case, we observe an alternative to the C-state as thermal noise can produce sufficient deviation from the C-state to let the system evolve to a state where most magnetizations point approximately parallel to the hard axis at large stresses.

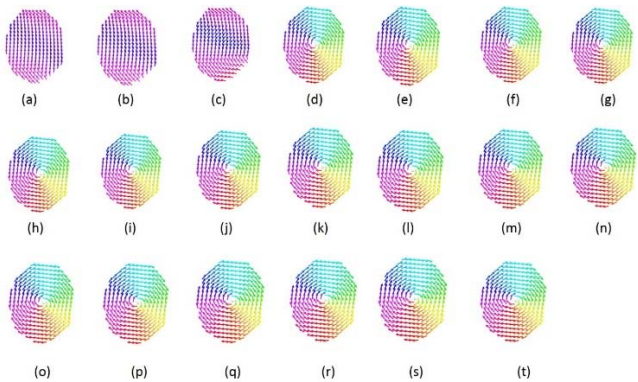

**Figure S6**: Evolution of magnetization at T=0 K with stress for a nanomagnet with dimensions of 200 nm × 160 nm × 12 nm, for the following stress scenarios: (a) 5 MPa, (b) 10 MPa, (c) 15 MPa, (d) 20 MPa, (e) 30 MPa, (f) 40 MPa, (g) 50 MPa, (h) 60 MPa, (i) 70 MPa, (j) 80 MPa, (k) 90 MPa, (l) 100 MPa. , (m) 110 MPa, (n) 120 MPa, (o) 130 MPa, (p) 140 MPa, (q) 150 MPa, (r) 160 MPa, (s) 170 MPa, (t) 180 MPa

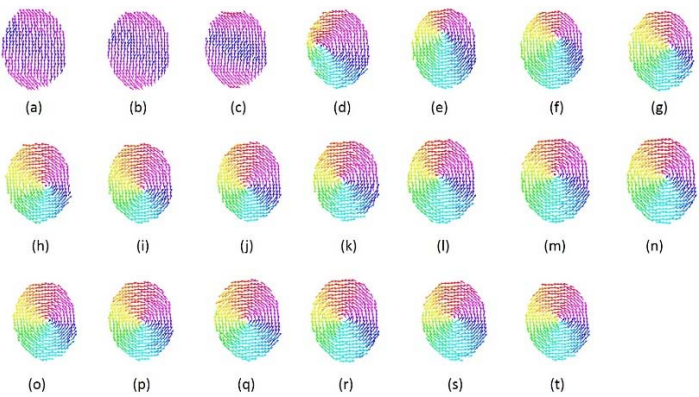

**Figure S7**: Evolution of magnetization at T=300 K with stress for a nanomagnet with dimensions of 200 nm × 160 nm × 12 nm, for the following stress scenarios: a) 5 MPa, (b) 10 MPa, (c) 15 MPa, (d) 20 MPa, (e) 30 MPa, (f) 40 MPa, (g) 50 MPa, (h) 60 MPa, (i) 70 MPa, (j) 80 MPa, (k) 90 MPa. Once again, the magnetization shows more deviation from the vortex state in Fig. 6 due to thermal noise.